\documentstyle[prl,aps,twocolumn]{revtex}

\newcommand{\beq}{\begin{equation}}
\newcommand{\eeq}{\end{equation}}
\newcommand{\bea}{\begin{eqnarray}}
\newcommand{\eea}{\end{eqnarray}}
\newcommand{\pri}{^{\prime}}

\newcommand{\bs}{{\bf{S}}}

\newcommand{\etal}{{\em et al.}}

\def\tit#1#2#3#4#5{{#1} {\bf #2}, #3 (#4).}

\def\prl{Phys. Rev. Lett.}

\def\prb{Phys. Rev. B}

\def\zpb{Z. Phys. B}

\begin{document}
\draft

\twocolumn[\hsize\textwidth\columnwidth\hsize\csname @twocolumnfalse\endcsname

\title{Properties of a classical spin liquid: the Heisenberg pyrochlore antiferromagnet}
\author{R. Moessner and J. T. Chalker}
\address{Theoretical Physics, Oxford University,
1 Keble Road, Oxford OX1 3NP, UK}
\maketitle
\begin{abstract}
We study the low-temperature behaviour of the classical Heisenberg 
antiferromagnet with nearest neighbour interactions on the pyrochlore
lattice. Because of geometrical frustration, 
the ground state of this model has an extensive number of degrees of freedom.
We show, by analysing the effects of small fluctuations around
the ground-state manifold, and from the results of Monte Carlo and 
molecular dynamics simulations, that the system is disordered 
at all temperatures, $T$, and has a finite relaxation time,
which varies as $T^{-1}$ for small $T$.
\end{abstract}


]

\input{psfig}

In recent years, geometrically frustrated antiferromagnets 
have been identified as a distinct class of materials, 
separate both 
from unfrustrated antiferromagnets
and from conventional spin-glasses \cite{reviews}.
Most characteristically, they remain in the paramagnetic phase,
down to a freezing temperature, $T_F$, which is small on the scale 
set by the interaction strength, as measured via the magnitude
of the Curie-Weiss constant, $\Theta_{CW}$.
This behaviour appears 
to be a consequence of 
their structures, with magnetic ions arranged in
corner-sharing frustrated units -- triangles or 
tetrahedra -- favouring high ground-state degeneracy. 

Compounds in this class include
$SCGO$ \cite{SCGO}, in which a proportion of the magnetic 
ions occupy the
sites of a kagome lattice, and the oxide \cite{gaurei,dunsiger} 
and flouride \cite{harrisneutron1,harrisneutron2} pyrochlores,
in which the magnetic ions form tetrahedra,
as illustrated in Fig.\ 1. 
Magnetic correlations in these materials,
determined from
neutron scattering \cite{SCGO,gaurei,harrisneutron1,harrisneutron2} 
and muon spin relaxation \cite{dunsiger,muon} measurements, are short-ranged,
with fluctuations that slow down as $T$ is 
reduced towards $T_F$ \cite{reviews}. 
\begin{figure}
\vspace{-1.7cm}
\centerline{\psfig{figure=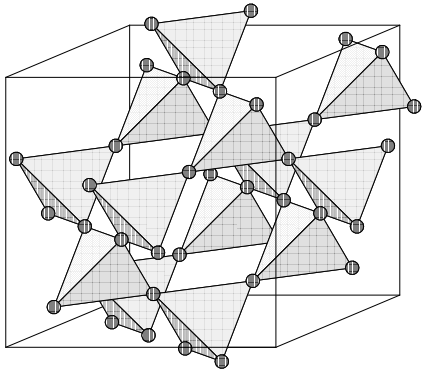,width=7cm}}
\vspace{-4cm}
\caption{The pyrochlore lattice}
\label{fig:pyrochlore}
\end{figure}
An important step towards a theory of geometrically frustrated 
antiferromagnets is to understand the behaviour of the classical
Heisenberg model defined with nearest neighbour interactions on 
the appropriate lattices. This simplified description may be sufficient
in the paramagnetic phase, and is a natural starting
point for the treatment of various additional features of real materials
(anisotropy, disorder, dipolar interactions and quantum fluctuations)
that might be relevant, especially below $T_F$.
For the lattices concerned, the
Heisenberg antiferromagnet has ground states with extensive numbers of
degrees of freedom. Properties in the 
temperature range $T \ll |\Theta_{CW}|$ are controlled by small amplitude
fluctuations around the ground-state manifold:
the free energy of fluctuations may select specific
ground states, a phenomenon known as {\em order by disorder} \cite{obd},
while the long-time dynamics results from coupling between these
fluctuations and the ground-state coordinates. Dynamical 
correlations, in particular, are potentially one of the most
interesting aspects of these systems, but have so far
received only limited attention \cite{reimersmc,keren}.

In this paper we analyse the low-temperature statistical mechanics
and dynamics of the classical pyrochlore Heisenberg antiferromagnet,
and place
our results in a broader setting.
Most importantly, we show that the system is, as
proposed in early work by Villain \cite{villain},
an example of a cooperative paramagnet or classical
spin liquid. It does not display order by disorder,
and at small $T$ the spin autocorrelation function
(with precessional dynamics) decays in time, $t$,
as $\langle {\bf S}_i(0).{\bf S}_i(t) \rangle = \exp(-cTt)$,
where $c$ is a constant. This behaviour is in striking contrast 
to that of the kagome Heisenberg antiferromagnet, previously
the best-studied example of geometric frustration, in which 
fluctuations
select coplanar spin configurations in the limit 
$T \to 0$ \cite{chalkerkagome}. 

We take, as a general starting point,
$n$-component classical spins, $\bs_i$, with $|\bs_i|=1$,
arranged in $q$-site, corner-sharing units:
the kagome and pyrochlore lattices have
$q=3$ and $q=4$, respectively \cite{footnote}.
An antiferromagnetic exchange interaction, of strength $J$, couples
each spin with its $2(q-1)$ nearest neighbours
in the two units to which it belongs, so that the Hamiltonian is
\bea
H & = &   
J \sum_{<i,j>} \bs_i \cdot \bs_j \label{eq:spinhamil} 
\equiv \frac{J}{2}\sum_{\alpha}|{\bf L}_{\alpha}|^2-~\frac{J}{2}Nq,
\label{eq:spinhamil2}
\eea 
where the sum on $\left<i,j \right>$\ runs over all 
neighbouring pairs, the sum on $\alpha$ runs over the $N$ units
making up the system, and
${\bf L}_{\alpha}$\ is total spin in unit $\alpha$.

The ground-state degeneracy can be demonstrated using a Maxwellian
counting argument, as follows \cite{reimers}.
From Eq.~\ref{eq:spinhamil2}, a configuration is a ground state provided 
${\bf L}_{\alpha}=0$ for each unit separately.
The system has in total $F=N(n-1)q/2$ degrees of freedom,
and the requirement that ${\bf L}_{\alpha}=0$ for all $\alpha$
imposes $K=Nn$ constraints. The ground-state manifold has $D=F-K$ dimensions 
if these constraints are independent.
Of the physically realisable cases ($q \leq 4$, $n \leq 3$), it is only
for $q=4$, $n=3$ that $F-K=N[n(q-2)-q]/2$ is positive and extensive
(taking the value $F-K=N$), and it is partly for this reason
that the pyrochlore Heisenberg antiferromagnet
is particularly interesting.
In other systems, an extensive ground-state dimension can
arise only if the constraints are not all independent,
as happens starting from coplanar spin configurations of 
the kagome Heisenberg model.

The ground-state manifold has two simple properties which should
have direct physical consequences. First,
the same counting argument shows, for a system in which
$F-K$ is positive and extensive, that there exist
degrees of freedom within the ground state which are strictly local.
That is, for a region of the system which is sufficiently large,
the ground state remains degenerate even if spins at the surface 
of the region are
held fixed, since $F-K$, proportional to volume,
is larger than the number of additional constraints, 
proportional to surface area.
It seems likely that such local modes will cause rapid relaxation of
spin correlations.
Second,
restricting attention to the pyrochlore lattice with $n=2$ or $3$
and open boundary conditions,
we have been able to 
prove that the manifold is connected.
The central idea is that any ground state can be 
deformed continuously into a reference
state, without leaving the manifold; details will be 
given elsewhere \cite{elsewhere}.
An implication is that the system does not have internal energy
barriers: if it were to freeze,
it could do so only 
because of dynamical bottle-necks or free-energy barriers.
 
At low but non-zero temperature, $0<T\ll J$, all accessible 
configurations lie close to the ground-state manifold,
and thermal fluctuations generate
a probability distribution for the system
on this manifold.
The question of whether the system shows order by disorder
is a question about the limiting form
of this distribution as $T \to 0$.
To specify the problem more precisely, let ${\bf x}$ denote coordinates
on the ground-state manifold, and let ${\bf y}$ be 
(locally defined) coordinates
in the remaining, orthogonal directions in configuration space.
The leading term in the energy, sufficient for $T \ll J$,
is quadratic in ${\bf y}$. Choosing axes
that diagonalise this quadratic form, we have
\bea
H \approx H_2 =\sum_l \epsilon_l({\bf x})y_l^2. 
\label{H}
\eea 
Retaining only quadratic terms in $H$ and
integrating over ${\bf y}$, the (unnormalised) ground-state 
probability distribution is
\bea
Z({\bf x})= \int d\{y_l\} e^{-\beta H_2} 
\propto \prod_l (k_{{\rm B}}T/\epsilon_l({\bf x}))^{1/2}.
\label{Z}
\eea
If $Z({\bf x})$, calculated in this way, is normalisable,
the system does not show order by disorder: instead, it
explores all ground states in the limit $T \to 0$, with a 
probability density proportional to $Z({\bf x})$.
Alternatively, 
for a particular ground state, ${\bf x}_0$, 
some of the $\epsilon_l({\bf x}_0)$ may
vanish. Then $Z({\bf x})$ will diverge as ${\bf x}$ approaches ${\bf x}_0$.
If any such divergences are non-integrable,
one should keep higher order terms from Eq.\ \ref{H}
when calculating $Z({\bf x})$. The result of doing so will be,
in the limit $T \to 0$, a
distribution concentrated exclusively
on the sub-set of ground states for which $Z({\bf x})$
is divergent: these are the configurations selected by 
thermal fluctuations.

To decide whether there is, in fact, 
order by disorder, it is necessary to know
the number, $M$, of $\epsilon_l$ that vanish, and the dimension, $S$, of
the subspace on which this happens. 
Close to this subspace, we separate
${\bf x}\equiv ({\bf u},{\bf v})$ into an 
$S$-dimensional component 
${\bf u}$, lying within it, and a 
$(D-S)$-dimensional component
${\bf v}$, locally orthogonal to it.
Then, transforming ${\bf v}$ into radial and angular variables, $v$,
and
\mbox{\boldmath $\Omega$}, we have, at small $v$, the behaviour
$\epsilon_l({\bf x}) \propto v^2$ for $M$ of the $\epsilon_l$'s.
Hence, $Z({\bf x})$ diverges as $v^{-M}$ for small $v$, 
at fixed 
\mbox{\boldmath $\Omega$} 
and ${\bf u}$, 
and
the potentially divergent contribution to its normalisation is
\bea
\int Z({\bf u},{\bf v}) d{\bf v}\, \propto \,\int v^{D-S-M-1} dv.
\eea
The integral is actually divergent at small $v$, and the system has order 
as $T \to 0$, if $D-S-M \leq 0$; 
otherwise, the subspace is not selected.

In principle, one should calculate the value of $D-S-M$ for all possible ordering patterns. In fact, we examine only the simplest candidates, 
and obtain for these only the extensive part of $D-S-M$,
checking our conclusions using Monte Carlo
simulations. Specifically, we
test for collinear spin order in tetrahedra, and coplanar order in 
triangles. We find \cite{elsewhere} that $D-S-M$ increases with $n$, passing through the
marginal value, zero, at $n=3$ in the first case, and $n=4$ in the second case.
Simulations of both marginal systems -- the Heisenberg model
on the pyrochlore lattice (Refs \cite{reimersmc,martin} and as described
below), and four-component spins on the kagome lattice \cite{rute} --
indicate that they are disordered.
We conclude that the only models of this kind which display
order by disorder are the $XY$-pyrochlore and Heisenberg kagome 
antiferromagnets.

We next present the results of Monte Carlo simulations
of pyrochlore antiferromagnets
with two- and three-component spins, which 
extend pioneering calculations by Reimers \cite{reimersmc}
and Zinkin \cite{martin},
and confirm the above conclusions.
We treat
systems of size varying 
from $2N=4$\ to $2N=19652$\ spins, over a temperature range 
extending down to $T/J=5\times 10^{-5}$. We use run lengths 
of order $10^6$ Monte Carlo steps per spin and check that the same 
results are obtained from both random and ordered initial configurations.

If order by disorder occurs, it leaves a signature in the
heat capacity per spin, $C$, of the classical model at low temperature,
which can be identified without advance knowledge of 
the ordering pattern \cite{chalkerkagome}. 
The value of $C$ reflects the nature of fluctuations in the system:
each coordinate $y_l$ for which $\epsilon_l$ is non-zero
contributes $k_{{\rm B}}/2$\  to the total heat capacity,
while coordinates for which $\epsilon_l=0$, so that the
associated energy varies as $y_l^4$, make contributions of $k_{{\rm B}}/4$.
Since there are in total $n/2$ coordinates $\{y_l\}$ per spin,
and since
collinear order introduces one such quartic mode per tetrahedron,
or half a mode per spin, we expect values for $C$, in units of
$k_{{\rm B}}$, of 
$n/4$ without order, and $(n/4 - 1/8)$ with 
collinear order.
At $T/J \sim 10^{-4}$, we find $C=0.376\pm 0.002$
for the $XY$ model, and $C=0.747\pm 0.002$
for the Heisenberg model.
For $XY$ spins, this confirms that there is one quartic mode per tetrahedron,
and hence ordering.
For Heisenberg spins, it sets an upper limit of 0.04 quartic
modes per tetrahedron: there is no order by disorder.

\begin{figure}
\vspace{-0.5cm}
\centerline{\psfig{figure=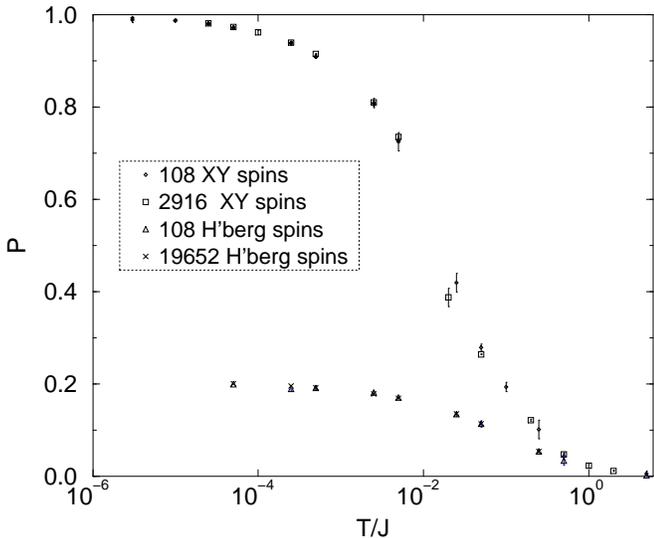,width=10cm}}
\vspace{-0.5cm}
\caption{The collinearity parameter, $P$, for $XY$ and Heisenberg
spins. Error bars are smaller than the symbols except where shown.} 
\label{fig:collxy}
\end{figure}

To show explicitly that ordering in the $XY$ model is collinear,
and that there is no such order in the Heisenberg model,
we introduce a measure of local collinearity
\bea
P\equiv 
\frac{1}{6N}\sum_{\left< i, j \right>}
\frac{n}{n-1}\left(\langle\left( \bs_i \cdot \bs_j\right)^2\rangle
-\frac{1}{n} \right),
\eea 
defined so that $P=1$ for collinear spins and
$P=0$ in the high-temperature limit.
Its temperature dependence is shown in Fig. \ref{fig:collxy}. 
For the $XY$ model, $P$ approaches 1 as $T \to 0$,
providing unequivocal evidence of order.
Deviations vary as
$1-P\propto (T/J)^{1/2}$, behaviour that
can be understood on the basis of the analysis of Ref.\ \cite{chalkerkagome}.
For Heisenberg spins, $P$ has a maximum value of $0.2$,
which is essentially independent of system size, and of temperature
in the range $10^{-4} < T/J < 10^{-3}$.
The fact that $P$ remains less than $1$ as $T\to 0$
is a demonstration that the Heisenberg model
does not
have collinear order induced by fluctuations.
The constant,  non-zero value of $P$ at the lowest temperatures
merely indicates that neighbouring spins
are on average more collinear in
ground states than in the high-temperature limit.

Finally, we turn to the low-temperature dynamics of the 
Heisenberg pyrochlore antiferromagnet. Since different
ground states are separated neither by energy barriers
(the ground-state manifold is connected), nor by large
free energy barriers (the system does not display order by disorder),
one might expect correlations to relax rapidly even at low temperature,
We find that they indeed do,
in the sense that the relaxation time 
increases at low temperature only as 
$T^{-1}$ and not exponentially.
The equation of motion is
\bea
\frac{d\bs_i}{dt}=
\bs_i\times {\bf H}_i(t)
\equiv -J\,\bs_i\times({\bf L}_{\alpha}+{\bf L}_{\beta}),
\label{eqnofmotion}
\eea
where ${\bf H}_i(t)$ is the exchange field acting at site $i$, which
can be expressed in terms of
${\bf L}_{\alpha}$ and ${\bf L}_{\beta}$, the total spins
of the two tetrahedra to which $\bs_i$ belongs.
Solving this equation in the harmonic
approximation, by
linearising around a ground state, yields $2N$
normal modes. If the ground state is a generic one,
in the sense that none of the $\epsilon_l$ of Eq.\ \ref{H}
are zero, then $3/4$ of these modes
have finite frequencies and the remaining
$1/4$ have zero frequency. The canonical coordinates
for the zero modes
represent displacements in phase-space
that lie within the ground-state manifold.
In order to study non-trivial aspects of the
long-time dynamics, it is of course
necessary to go beyond the harmonic approximation.
At low temperature, anharmonic effects are small
and there is a separation of time-scales: the periods of finite-frequency
modes are temperature-independent, and the shortest of these
sets a scale  of order $J^{-1}$, 
while the long time-scale for motion around the ground-state manifold
increases as $T$ decreases. 
For short times, the exchange field, ${\bf H}_i(t)$ consists simply of
oscillatory contributions from each of the finite-frequency modes.
Over long times, the amplitudes of these modes vary, 
because of anharmonic coupling, and ${\bf H}_i(t)$ 
is a randomly fluctuating function. Its relevant properties
on these scales are its mean,
$\langle {\bf H}_i(t) \rangle =0$, and its low-frequency spectral density
\bea
\int_{-\infty}^{\infty}dt' \langle {\bf H}_i(t) \cdot {\bf H}_i(t')  \rangle\, 
\equiv\, 2\Gamma\,.
\label{D}
\eea
Noting that: (i) $\langle|{\bf H}_i(t)|^2\rangle \sim J^2 
\langle |{\bf L}_{\alpha}|^2 \rangle \sim Jk_{\rm B}T$; and (ii)
only the lowest frequency modes contribute to the time-integral in 
Eq.\ \ref{D}, we find \cite{elsewhere}
$\Gamma \propto Jk_{\rm B}T\rho(0^+)$, where $\rho(\omega)$ is the 
density in frequency, $\omega$, of the finite-frequency modes.
We have checked, by diagonalising the linearised equations
of motion numerically, that $\rho(0^+)$  is non-zero; we conclude that
$\Gamma=cT$ where $c$ is a constant.
We can therefore
calculate long-time spin correlations from Eq.\ \ref{eqnofmotion}, by
treating ${\bf H}_i(t)$ as Gaussian white noise with the correlator
$\langle {\bf H}_i(t) \cdot {\bf H}_i(t')  \rangle = 2\Gamma \delta(t-t')$.
Solving the resulting Langevin equation, we obtain
\bea
\langle \bs_i(0) \cdot \bs_i(t) \rangle = e^{-cTt}.
\label{auto}
\eea

To test these ideas, we have carried out molecular dynamics simulations
in which we compute
the spin autocorrelation function $A(t)\equiv\left<
\bs(t\pri)\cdot\bs(t+t\pri)\right>$.
Similar calculations for the kagome Heisenberg antiferromagnet
have been described by Keren \onlinecite{keren}.
A Monte Carlo simulation is used
to generate uncorrelated, thermalised initial configurations, from which the equation of motion, Eq.\ \ref{eqnofmotion}, is integrated
using a fourth-order Runge-Kutta algorithm.
The integration time-step is
chosen so that energy is conserved to at least one part in $10^{8}$.
Finite size effects are suppressed by constraining the
total spin of the entire system to be near zero.

\begin{figure}
\vspace{-1cm}
\centerline{\psfig{figure=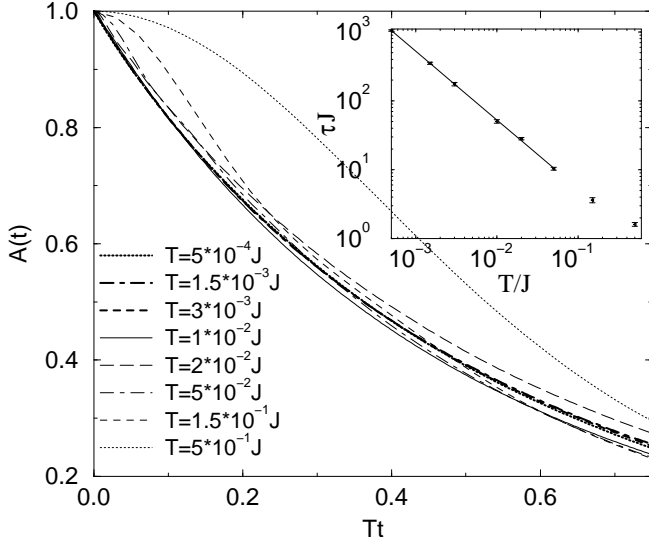,width=10cm}}
\vspace{-0.5cm}
\caption{The autocorrelation function as a function of the rescaled
time $t\times T$ over one and a half decay times. Inset: the decay
time, $\tau$, as a function of temperature.
}
\label{fig:zerop}
\end{figure}

We expect at low temperature, from Eq.\ \ref{auto}, 
that $A(t)$ should depend on $T$ and $t$ only through the scaling variable
$Tt$. $A(t)$\ is shown in Fig.\ \ref{fig:zerop} as a function of the
rescaled time, $Tt$, at eight different temperatures in the range
$5\times 10^{-4} \leq T/J \leq 0.5$,\, for a system of size 2048 spins.
There are no adjustable parameters in the
construction of this plot. Except at the highest temperatures,
the data collapse onto a single curve, which is exponential to the precision
of our calculations. 
To examine quantitatively the accuracy of the temperature scaling, we 
extract a decay time, $\tau$, from $A(t)$ at each temperature $T/J<0.15$,
and fit the temperature-dependence of $\tau$ to the power law
$\tau \propto T^{-\zeta}$  (Fig.\ \ref{fig:zerop} inset), obtaining
$\zeta=0.998 \pm 0.012$.
These results agree with and confirm our predictions.

Some experimental properties of 
pyrochlore antiferromagnets are consistent with
our findings, although a quantitative comparison is not possible.
Inelastic neutron scattering from $CsNiCrF_6$ \cite{harrisneutron2}
fits the response expected for spin diffusion, with a relaxation time 
that decreases with $T$, and the spin dynamics of
oxide pyrochlores, observed in $\mu SR$  \cite{dunsiger},
also slow down as $T$ is reduced. 
We find, in agreement with Reimers \cite{reimersmc},
that the freezing transition observed in all 
such materials at the lowest
temperatures is absent from the Heisenberg
model. 

In summary, we have presented a description of the low-temperature properties
of the
Heisenberg pyrochlore antiferromagnet. We have shown that the system
does not order,
and that its low-temperature
dynamics is diffusive on a timescale set by the inverse temperature. 

We thank P. Chandra, R. Cowley, M. Harris, P. Holdsworth
and M. Zinkin for helpful discussions. We are also 
grateful to 
the Institute for Theoretical Physics, UCSB, 
for hospitality while this work was completed. It was supported in
part through EPSRC Grant GR/J8327 and NSF Grant PHY94-07194.

\end{document}